\documentclass[prd,aps,showpacs,preprintnumbers,amssymb]{revtex4}
\usepackage{axodraw}
\usepackage{color}
\usepackage{epsf}

\def\e3p{$\eta \rightarrow 3 \pi$}

\begin{document}
\title{%
\hfill{\normalsize\vbox{%
\hbox{}
 }}\\
{A  semi perturbative method for QED}}

\author{Renata Jora
$^{\it \bf a}$~\footnote[2]{Email:
 rjora@theory.nipne.ro}}

\author{Joseph Schechter
 $^{\it \bf b}$~\footnote[4]{Email:
 schechte@phy.syr.edu}}

\affiliation{$^{\bf \it b}$ National Institute of Physics and Nuclear Engineering PO Box MG-6, Bucharest-Magurele, Romania}

\affiliation{$^ {\bf \it c}$ Department of Physics,
 Syracuse University, Syracuse, NY 13244-1130, USA}

\date{\today}

\begin{abstract}
We compute the QED beta function using a new method of functional integration. It turns out that in this procedure the beta function contains only the first two orders coefficients
and thus corresponds to  a new renormalization scheme, long time supposed to exist.
\end{abstract}
\maketitle

\section{Introduction}

Beta functions in the standard renormalization procedures are known for QED up to the fifth order whereas for QCD up to the fourth one \cite{Vladimirov}-\cite{Baikov}.
In \cite{Jora} we introduced a new method of functional integration and obtained an all order correction to the scalar mass for the $\Phi^4$ theory with a single scalar and without spontaneous symmetry breaking. We will use these findings as laboratory for studying more complex theories; in the present work we shall apply the method illustrated in \cite{Jora}
to the case of QED with  fermions in the fundamental representation. This is obviously  a step forward as this model displays more complicated interactions and set-up. We will postpone the discussion of in what measure the method can be implemented for the more realistic case of the standard model for further work.
We start with the Lagrangian,
\begin{eqnarray}
{\cal L}_{QED}=\bar{\Psi}(i\gamma^{\mu}D_{\mu}-m)\Psi-\frac{1}{4}F_{\mu\nu}F^{\mu\nu}.
\label{lagr56}
\end{eqnarray}

We rewrite the Lagrangian in Eq. (\ref{lagr56}) in terms of the Fourier modes because we would like to integrate over these in the functional approach:
\begin{eqnarray}
&&\int d^4 x{\cal L}_{QED}=\frac{1}{2}\int \frac{d^4k}{(2\pi)^4}A_{\mu}(k)[-k^2g^{\mu\nu}+(1-\frac{1}{\xi})k^{\mu}k^{\nu}]A_{\nu}(-k)+
\nonumber\\
&&+\int \frac{d^4k}{2\pi)^4}[\Psi(k)(\gamma^{\mu}k_{\mu}-m)\Psi(k)]-e\int \frac{d^4k}{(2\pi)^4}\frac{d^4p}{(2\pi)^4}\bar{\Psi}(p)\gamma^{\mu}\Psi(-p+k)A_{\nu}(-k),
\label{dec4567}
\end{eqnarray}
where we count over both positive and negative k modes.

We shall integrate expression in (\ref{dec4567}) following the method introduced in \cite{Jora}.
We consider the function of the Fourier modes of the gauge field as a quadratic form plus a linear term. By forming the corresponding gaussian form we integrate over the gauge fields to
obtain:
\begin{eqnarray}
&&W[0]=\int d\Psi d\bar{\Psi} d A_{\mu}\exp[i\int d^4 x {\cal L}_{QED}]={\rm const}
\int d\Psi d\bar{\Psi}(\det[k^2g^{\mu\nu}+(1-\frac{1}{\xi})k^{\mu}k^{\nu}])^{-1/2}\times
\nonumber\\
&&\exp\left[i[\int \frac{d^4k}{(2\pi)^4} [\bar{\Psi}(k)(\gamma^{\mu}k_{\mu}-m)\Psi(k)]-\int e^2\frac{d^4k}{(2\pi)^4} \frac{d^4q}{(2\pi)^4} \frac{d^4p}{(2\pi)^4} [\frac{1}{4}\bar{\Psi}(p)\gamma^{\mu}\Psi(p+k)D_{\mu\nu}^{-1}\bar{\Psi}(q)\gamma^{\nu}\Psi(q-k)]\right]
\label{rez4356}
\end{eqnarray}
Here we denoted:
\begin{eqnarray}
D^{\mu\nu}=-k^2g^{\mu\nu}+(1-\frac{1}{\xi})k^{\mu}k^{\nu},
\label{not678}
\end{eqnarray}
and $\xi$ is the usual gauge parameter. Moreover we counted only over the modes with $k_0>0$ case in which the kinetic term for the gauge field appears without the factor of $\frac{1}{2}$ in front.
We shall integrate in Eq. (\ref{rez4356}) by introducing  a new variable  $\eta_{\mu}$ and a delta function:
\begin{eqnarray}
&&W[0]={\rm const}\int d\bar{\Psi}d \Psi d\eta_{\mu}(\det[k^2g^{\mu\nu}+(1-\xi)k^{\mu}k^{\nu}])^{-1/2}\delta(\eta_{\mu}-\bar{\Psi}\gamma^{\mu}\Psi)\times
\nonumber\\
&&\exp[i[\int \frac{d^4k}{(2\pi)^4} \bar{\psi}(k)(\gamma^{\mu}k_{\mu}-m)\Psi(-k)-\int\frac{e^2}{4}\frac{d^4k}{(2\pi)^4}\eta^{\mu}D_{\mu\nu}^{-1}\eta^{\nu}]]
\label{subst6578}
\end{eqnarray}

We further express the delta function in terms of its exponential representation to get:
\begin{eqnarray}
&&W[0]={\rm const}\int d\bar{\Psi}d \Psi d\eta_{\mu} d K_{\mu}(\det[-i(k^2g^{\mu\nu}+(1-\frac{1}{\xi})k^{\mu}k^{\nu})])^{-1/2}\exp[i\int d^4 x K_{\mu}(\eta_{\mu}-\bar{\Psi}\gamma^{\mu}\Psi)]\times
\nonumber\\
&&\exp[i[\int \frac{d^4k}{(2\pi)^4} \bar{\Psi}(k)(\gamma^{\mu}k_{\mu}-m)\Psi(-k)-\int\frac{e^2}{4}\frac{d^4k}{(2\pi)^4}\eta^{\mu}(k)D_{\mu\nu}^{-1}\eta^{\nu}(-k)]]
\label{s566578}
\end{eqnarray}

We then integrate over the $\eta_{\mu}$ field by forming quadratic forms out of the expression in the exponent. This leads to:
\begin{eqnarray}
&&W[0]={\rm const}\int d\bar{\Psi}d \Psi  d K_{\mu}(\det[-i(k^2g^{\mu\nu}+(1-\xi)k^{\mu}k^{\nu})])^{-1/2}(\det[\frac{1}{e^2}D_{\mu\nu}])^{1/2}\times
\nonumber\\
&&\exp[i\int \frac{d^4k}{(2\pi)^4}\frac{1}{2e^2}K_{\mu}(k)D^{\mu\nu}(k)K_{\nu}(-k)]\times
\nonumber\\
&&\exp[i\int \frac{d^4k}{(2\pi)^4}\bar{\Psi}(k)(\gamma^{\mu}k_{\mu}-m)\Psi(k)-i\int \frac{d^4k}{(2\pi)^4}\frac{d^4p}{(2\pi)^4}K_{\mu}(k)\bar{\Psi}(p)\gamma^{\mu}\Psi(-k+p)]=
\nonumber\\
&&={\rm const}\int  d K_{\mu}(\det[\frac{1}{2}(k^2g^{\mu\nu}+(1-\xi)k^{\mu}k^{\nu})])^{-1/2}(\det[\frac{2i}{e^2}D_{\mu\nu}])^{1/2}\times
\nonumber\\
&&\exp[i\int \frac{d^4k}{(2\pi)^4}\frac{1}{2e^2}K_{\mu}(k)D^{\mu\nu}(k)K_{\nu}(-k)]\times\det[(\gamma^{\mu}k_{\mu}-m)\delta_{m,n}-\gamma^{\mu}(K_{\mu})_{-m-n=k}]
\label{finrez6578}
\end{eqnarray}

Note that the result in Eq. (\ref{finrez6578}) is exactly that of QED with the field $A_{\mu}$ replaced by the new variable $K_{\mu}$. It seems that our derivation although correct is redundant. However in the next sections we will show that this procedure in its intermediate steps allows us to extract the corrections to the beta function for the electric charge in a new simplified renormalization scheme.

\section{The two point function}

We write the expression for the two point function in the Fourier space:
\begin{eqnarray}
\int d A_{\mu} d\bar{\Psi}d \Psi A_{\rho}(p)A_{\sigma}(q)\exp[i\int d^4 x{\cal L}_{QED}]
\label{rez54678}
\end{eqnarray}

We perform the same change of variable as in the first section $A_{\nu}(k)\rightarrow A_{\nu}-\frac{e}{2}\Psi(p)\frac{\gamma^{\nu}}{D_{\mu\nu}}\Psi(p-k)$ to obtain:
\begin{eqnarray}
&&I_{\rho\sigma}=\int d A_{\mu} d\bar{\Psi}d \Psi [A_{\rho}(p)A_{\sigma}(q)+\frac{e^2}{4}\bar{\Psi}(r)\frac{\gamma^{\mu}}{D_{\mu\rho}(p)}\Psi(r+p)\bar{\Psi}(u)
\frac{\gamma^{\nu}}{D_{\nu\sigma}(q)}\Psi(u+q)]\times
\nonumber\\
&&\exp[i\frac{1}{2}\int \frac{d^4k}{(2\pi)^4}A_{\mu}(k)D_{\mu\nu}A_{\nu}(-k)+\int \frac{d^4k}{(2\pi)^4} \bar{\Psi}(k)(\gamma^{\mu}k_{\mu}-m)\Psi(k)-
\nonumber\\
&&\int e^2\frac{d^4k}{(2\pi)^4} \frac{d^4q}{(2\pi)^4} \frac{d^4p}{(2\pi)^4} [\frac{1}{4}\bar{\Psi}(p)\gamma^{\mu}\Psi(p+k)D_{\mu\nu}^{-1}\bar{\Psi}(q)\gamma^{\nu}\Psi(q-k)]],
\label{rez67589}
\end{eqnarray}
where this time the variable $A_{\mu}$ is the new variable and we dropped the odd terms that lead to zero in the functional integration.

Since the integrals over the gauge and fermion  fields are independent one can write:

\begin{eqnarray}
&&I_{\rho\sigma}=\int d A_{\mu}  [A_{\rho}(p)A_{\sigma}(q)]\exp[i\frac{1}{2}\int \frac{d^4k}{(2\pi)^4}A_{\mu}(k)D_{\mu\nu}A_{\nu}(-k)]\times
\nonumber\\
&&\int d\bar{\Psi}d \Psi\exp[i\int \frac{d^4k}{(2\pi)^4} \bar{\Psi}(k)(\gamma^{\mu}k_{\mu}-m)\Psi(k)-
\int e^2\frac{d^4k}{(2\pi)^4} \frac{d^4q}{(2\pi)^4} \frac{d^4p}{(2\pi)^4} [\frac{1}{2}\bar{\Psi}(p)\gamma^{\mu}\Psi(p+k)D_{\mu\nu}^{-1}\bar{\Psi}(q)\gamma^{\nu}\Psi(q-k)]]+
\nonumber\\
&&\int d A_{\mu}  \exp[i\frac{1}{2}\int \frac{d^4k}{(2\pi)^4}A_{\mu}(k)D_{\mu\nu}A_{\nu}(-k)]\times
\nonumber\\
&&\int d\bar{\Psi}d \Psi [\frac{e^2}{4}\bar{\Psi}(r)\frac{\gamma^{\mu}}{D_{\mu\rho}(p)}\Psi(r+p)\bar{\Psi}(u)
\frac{\gamma^{\nu}}{D_{\nu\sigma}(q)}\Psi(u+q)]\exp[i\int \frac{d^4k}{(2\pi)^4} \bar{\Psi}(k)(\gamma^{\mu}k_{\mu}-m)\Psi(k)-
\nonumber\\
&&\int e^2\frac{d^4k}{(2\pi)^4} \frac{d^4q}{(2\pi)^4} \frac{d^4p}{(2\pi)^4} [\frac{1}{2}\bar{\Psi}(p)\gamma^{\mu}\Psi(p+k)D_{\mu\nu}^{-1}\bar{\Psi}(q)\gamma^{\nu}\Psi(q-k)]]
\label{fullexp90}
\end{eqnarray}

Then the first term in Eq. (\ref{fullexp90}) can be separated and leads to upon functional integration:
\begin{eqnarray}
&&\int d A_{\mu}  [A^{\rho}(p)A^{\sigma}(q)]\exp[i\frac{1}{2}\int \frac{d^4k}{(2\pi)^4}A_{\mu}(k)D_{\mu\nu}A_{\nu}(-k)]=
\nonumber\\
&&\frac{i\xi}{2}\frac{\delta^2}{\delta k_{\rho} \delta k_{\sigma}}(\det[i(-k^2g^{\mu\nu}+(1-\frac{1}{\xi})k^{\mu}k^{\nu}])^{-1}=
\nonumber\\
&&\frac{-i}{k^2}(g^{\rho\sigma}-\frac{k^{\rho}k^{\sigma}}{k^2})(\det[i(-k^2g^{\mu\nu}+(1-\frac{1}{\xi})k^{\mu}k^{\nu}])^{-1},
\label{propf}
\end{eqnarray}
which corresponds to the free field propagator for the gauge field in the Landau gauge.

Eq. (\ref{propf}) needs further clarifications. First we note that quadratic kinetic operator that appears in the Lagrangian is singular such that we need to introduce the gauge parameter for consistency.  We shall conduct our calculations in specific gauge with $\chi=0$. First note that,
\begin{eqnarray}
i\xi\frac{\partial^2}{\partial k_{\rho} k_{\sigma}}i(-k^2A_{\mu}A_{\mu}+(1-\frac{1}{\xi})k^{\mu}k^{\nu}A_{\mu}A_{\nu})=
\xi[-2g^{\rho\sigma}A_{\mu}A^{\mu}+2(1-\frac{1}{\xi})A^{\rho}A^{\sigma}]=A^{\rho}A^{\sigma}
\label{res7689}
\end{eqnarray}
in the limit $\xi=0$. Thus our operator is adjusted for the Landau gauge and for every chosen gauge parameter one should associate a different operator.

In the same context let us note that our operator $D_{\mu\nu}$ satisfies the equation:

\begin{eqnarray}
(-k^2g^{\mu\nu}+(1-\frac{1}{\xi})k^{\mu}k^{\nu})\tilde{D}^{\nu\rho}=i\delta_{\mu}^{\rho},
\label{rel8796}
\end{eqnarray}
where $\tilde{D}^{\nu\rho}$ is the inverse operator:
\begin{eqnarray}
\tilde{D}^{\nu\rho}=\frac{-i}{k^2}(g^{\nu\rho}-(1-\xi)\frac{k^{\nu}k^{\rho}}{k^2}).
\label{inop789}
\end{eqnarray}

Then one can infer:
\begin{eqnarray}
\frac{\xi}{2}\frac{\delta^2}{\delta k^{\rho}\delta k^{\sigma}}\frac{1}{D_{\mu\nu}}=
\frac{1}{D_{\mu\sigma}D_{\nu\rho}}.
 \label{ident768950}
\end{eqnarray}
again in the same Landau gauge (Eq. (\ref{ident768950}) is not universal as it can be checked easily). Here the  derivative contains also other terms that we ignore because do not correpsond to one particle irreducible diagrams.

We then claim that:
\begin{eqnarray}
&&\int d\bar{\Psi}d \Psi[\frac{e^2}{4}\bar{\Psi}(r)\frac{\gamma^{\mu}}{D_{\mu\rho}(p)}\Psi(r+p)\bar{\Psi}(s)
\frac{\gamma^{\nu}}{D_{\nu\sigma}(q)}\Psi(s+q)]\times
\nonumber\\
&&\exp[i\int \frac{d^4k}{(2\pi)^4} \bar{\Psi}(k)(\gamma^{\mu}k_{\mu}-m)\Psi(k)-
\int e^2\frac{d^4k}{(2\pi)^4} \frac{d^4q}{(2\pi)^4} \frac{d^4p}{(2\pi)^4} [\frac{1}{4}\bar{\Psi}(p)\gamma^{\mu}\Psi(p+k)D_{\mu\nu}^{-1}\bar{\Psi}(q)\gamma^{\nu}\Psi(q-k)]]=
\nonumber\\
&&\frac{i\xi}{2}\frac{\partial^2}{\partial k^{\rho}k^{\sigma}}\int d\bar{\Psi}d \Psi \exp[i\int \frac{d^4k}{(2\pi)^4} \bar{\Psi}(k)(\gamma^{\mu}k_{\mu}-m)\Psi(-k)-
\nonumber\\
&&\int e^2\frac{d^4k}{(2\pi)^4} \frac{d^4q}{(2\pi)^4} \frac{d^4p}{(2\pi)^4} [\frac{1}{2}\bar{\Psi}(p)\gamma^{\mu}\Psi(p+k)D_{\mu\nu}^{-1}\bar{\Psi}(q)\gamma^{\nu}\Psi(q-k)]]
\label{ferm6758}
\end{eqnarray}

Now if we trace back from Eq. (\ref{finrez6578}) the contribution from Eq. (\ref{ferm6758}) we see that we need to consider the quantity:
\begin{eqnarray}
&&\frac{i\xi}{2}\frac{\partial^2}{\partial k ^{\rho} \partial k^{\sigma}} {\rm const}
\int d K_{\mu}(\det[\frac{2i}{e^2}D_{\mu\nu}])^{1/2}(\det[iD_{\mu\nu}])^{-1/2}\times
\nonumber\\
&&\exp[i\int \frac{d^4k}{(2\pi)^4}\frac{1}{2e^2}K_{\mu}(k)D^{\mu\nu}K_{\nu}(-k)]
\det[(\gamma^{\mu}p_{\mu}-m)\delta_{mn}-\gamma^{\mu}(K_{\mu})_{-m-n=k}]=
\nonumber\\
&&(\det[\frac{2}{e^2}])^{1/2}\frac{i\xi}{2}\frac{\partial^2}{\partial k^{\rho} \partial k^{\sigma}}\exp[i\int \frac{d^4k}{(2\pi)^4}\frac{1}{2e^2}K_{\mu}(k)D^{\mu\nu}K_{\nu}(-k)]
\det[(\gamma^{\mu}p_{\mu}-m)\delta_{mn}-\gamma^{\mu}(K_{\mu})_{-m-n=k}],
\label{partres678}
\end{eqnarray}
as it is evident that the two factors cancel each other.  We then make the change of variable $\frac{K_{\mu}}{e}\rightarrow K_{\mu}$ to obtain:
\begin{eqnarray}
I_{\rho\sigma}={\rm const} \frac{i\xi}{2}\frac{\partial^2}{\partial k^{\rho} \partial k^{\sigma}}\int d K_{\mu}
\exp[i\int \frac{d^4k}{(2\pi)^4}\frac{1}{2}K_{\mu}(k)D^{\mu\nu}K_{\nu}(-k)\det[(\gamma^{\mu}p_{\mu}-m)\delta_{mn}-e\gamma^{\mu}(K_{\mu})_{-m-n=k}]
\label{int657}
\end{eqnarray}

In order to extract  the corrections to the charge renormalization we need to make a new change of variable. The operator,
\begin{eqnarray}
O_{\alpha\mu}=\sqrt{\frac{i}{2k^2}}(k^2g^{\mu\alpha}-(1-\frac{1}{\sqrt{\xi}})k^{\mu}k^{\alpha})
\label{operat567}
\end{eqnarray}
satisfies the relation,
\begin{eqnarray}
O_{\alpha\mu}O^{\alpha}_{\nu}=-\frac{i}{2}D_{\mu\nu}
\label{form789}
\end{eqnarray}
where $D_{\mu\nu}$ is given in Eq. (\ref{not678}).
We then further make a new change of variable:
\begin{eqnarray}
K_{\mu}^{\prime}=K_{\alpha}O^{\alpha}_{\mu}.
\label{ch66878}
\end{eqnarray}
For simplicity we rename the new variable also $K_{\mu}$.  This leads to :

\begin{eqnarray}
&&I_{\rho\sigma}={\rm const} \frac{i\xi}{2}\frac{\partial^2}{\partial k^{\rho} \partial k^{\sigma}}
\int d K_{\mu}(\det[i(k^2g^{\mu\nu}-k^{\mu}k^{\nu}])^{-1}\exp[-\int \frac{d^4k}{(\pi)^4}\frac{1}{2}K_{\mu}(k)K_{\nu}(-k)]\times
\nonumber\\
&&\det[(\gamma^{\mu}p_{\mu}-m)\delta_{mn}-\gamma^{\mu}(K_{\alpha})_{-m-n=k}(O^{\alpha}_{\mu})^{-1}].
\label{intermedi890}
\end{eqnarray}

The first contribution is coming from:
\begin{eqnarray}
\frac{i\xi}{2}\frac{\partial^2}{\partial k^{\rho} \partial k^{\sigma}}\det[k^2g^{\mu\nu}-(1-\frac{1}{\xi})k^{\mu}k^{\nu}])^{-1}=
-i\frac{1}{k^2}(g^{\rho\sigma}-(1-\xi)\frac{k^{\rho}k^{\sigma}}{k^2})\det[i(k^2g^{\nu\mu}-k^{\mu\nu})])^{-1}
\label{rez5467}
\end{eqnarray}
which is  the free field propagator. Note that this is actually the contribution that we had obtained in Eq. (\ref{propf}) as we applied again the operator
$\frac{i\xi}{2}\frac{\partial^2}{\partial k^{\rho} \partial k^{\sigma}}$ to the full partition function.

Then the first extra contributions come from term of the type:
\begin{eqnarray}
\frac{\partial}{\partial k^{\rho}}(\det[k^2g^{\mu\nu}-(1-\frac{1}{\xi})k^{\mu}k^{\nu}])^{-1}\frac{\partial}{\partial k^{\sigma}}\det[(\gamma^{\mu}p_{\mu}-m)\delta_{mn}-e\gamma^{\mu}(K_{\alpha})_{-m-n=k}(O^{\alpha}_{\mu})^{-1}],
\label{next5674}
\end{eqnarray}
which we claim is zero. To prove that let us consider the first  factor in Eq. (\ref{next5674}):
\begin{eqnarray}
\frac{\delta}{\delta k^{\rho}}\det[k^2g^{\mu\nu}-(1-\frac{1}{\xi})k^{\mu}k^{\nu}]=
-8k^{\rho}\det[k^2g^{\mu\nu}-(1-\frac{1}{\xi})k^{\mu}k^{\nu}].
\label{rez678999}
\end{eqnarray}
In order to make the point we first outline our next steps. These comprise the calculation of the derivatives of the various $K_{\mu}$ terms. After we perform these we will change again the variable to $K_{\alpha}\rightarrow -2i O_{\alpha}^{\mu}K_{\mu}$. It turns out that one derivative of the type $\frac{\delta}{\delta k^{\rho}}$ applied to a $K_{\mu}$ term and after the new change of variable mentioned leads to terms which contain at most the inverse power $\frac{1}{\sqrt{\xi}}$. Then the result in Eq. (\ref{rez678999}) together with the result of differentiating the $K_{\mu}$ term will contain at most terms of type $\frac{1}{\sqrt{\xi}}$ which multiplied by $\xi$ in the limit $\xi=0$ leads to zero.

Second reason why we should not consider the term in Eq. (\ref{next5674}) is that this represents some first order correction. We are interested in computing in this approach the beta function and for all purposes we can consider for that:
\begin{eqnarray}
-i g_{\mu\nu}\frac{1}{k^2(1-\Pi(k^2))}.
\label{corect4577}
\end{eqnarray}
such that the terms proportional to $k^{\rho}k^{\sigma}$ can be neglected.  We shall do that for the rest of calculations for the sake of simplicity.

Moreover since we are actually interested in computing the beta function  we need $\Pi(0)$ so we can safely take the limit $k^2=0$ in the correction. In this context the variables $K_{\mu}(p)$ with $p^2\neq k^2$ that appear in the determinant can be considered as having a zero contribution and thus be neglected.

Next we need to determine the contributions proportional to:
\begin{eqnarray}
\frac{i\xi}{2}\frac{\partial^2}{\partial k^{\rho}\partial k^{\sigma}}\det[(\gamma^{\mu}p_{\mu}-m)\delta_{mn}-e\gamma^{\mu}(K_{\mu})_{-m-n=k}\frac{\sqrt{k^2}}{k^2g^{\mu\nu}-k^{\mu}k^{\nu}}]
\label{nexterm4567}
\end{eqnarray}

We need to consider the formula of differentiation of a determinant.
\begin{eqnarray}
&&\frac{1}{2}\frac{\partial^2}{\partial k^{\rho}\partial k^{\sigma}}\det A=
\frac{1}{2}\det A {\rm Tr}[\frac{\partial A}{\partial k^{\rho}}A^{-1}]{\rm Tr}[\frac{\partial A}{\partial k^{\sigma}}A^{-1}]+
\nonumber\\
&&\frac{1}{2}\det A{\rm Tr}[\frac{\partial^2 A}{\partial k^{\rho} \partial k^{\sigma}}A^{-1}]-\frac{1}{2}
\det A {\rm Tr}[\frac{\partial A}{\partial k^{\rho}}A^{-1}\frac{\partial A}{\partial k^{\sigma}}A^{-1}].
\label{det4567}
\end{eqnarray}
where,
\begin{eqnarray}
A=[(\gamma^{\mu}p_{\mu}-m)\delta_{mn}-e\gamma^{\mu}(K_{\mu})_{-m-n=k}\frac{\sqrt{k^2}}{k^2g^{\mu\nu}-k^{\mu}k^{\nu}}].
\label{not7689}
\end{eqnarray}

Here we should note that the term that contains $\frac{\partial^2 A}{\partial k^{\rho}k^{\sigma}}$ is zero by the same reasons as the term in Eq. (\ref{next5674}).

\section{The beta function}

In order to determine the charge renormalization given  by $\frac{1}{1-\Pi(0)}$ and in consequence the beta function in this method of functional integration we need to determine two terms:
\begin{eqnarray}
&&-i\xi\frac{1}{2}\det A {\rm Tr}[\frac{\partial A}{\partial k^{\rho}}A^{-1}\frac{\partial A}{\partial k^{\sigma}}A^{-1}]
\nonumber\\
&&i\xi\frac{1}{2}\det A{\rm Tr}[\frac{\partial A}{\partial k^{\rho}}A^{-1}]{\rm Tr}[\frac{\partial A}{\partial k^{\sigma}}A^{-1}],
\label{rez6789}
\end{eqnarray}
where the first term corresponds to the one loop contribution and the second to the two loops one. The beta function stops at two loops.

In order to determine terms of the type $\frac{\delta A}{\delta k^{\rho}}$ we note that the presence of the factor $\xi$ in front  means that we need to compute only the contribution proportional to $\frac{1}{\xi}$. Thus,
\begin{eqnarray}
\frac{\delta A}{\delta k_{\rho}}=e K_{\alpha}\sqrt{\frac{2k^2}{i}}(1-\frac{1}{\sqrt{\xi}})(g^{\mu\rho}\frac{k^{\alpha}}{k^2}+g^{\alpha\rho}\frac{k^{\mu}}{k^2})
\label{deriv54678}
\end{eqnarray}
Then we make the new change of variable: $K^{\alpha}=K_{\nu}^{\prime}\frac{k^2g^{\nu\alpha}-(1-\frac{1}{\xi})k^{\nu}k^{\alpha}}{\sqrt{k^2}}$ to find:
\begin{eqnarray}
e K_{\alpha}\frac{2k^2}{i}(1-\sqrt{\xi})(g^{\mu\rho}\frac{k^{\alpha}}{k^2}+g^{\alpha\rho}\frac{k^{\mu}}{k^2})\rightarrow e \sqrt{\frac{2k^2}{i}}\frac{1}{k^2}\frac{1}{\sqrt{\xi}}
K_{\nu}^{\prime}k^{\nu}g^{\mu\rho}
\label{contr5467}
\end{eqnarray}

We will show in some detail only how the first contribution in Eq. (\ref{rez6789}) is obtained. Thus,
\begin{eqnarray}
&&\int d K_{\mu} -i\xi\frac{1}{2}\det A {\rm Tr}[\frac{\partial A}{\partial k^{\rho}}A^{-1}\frac{\partial A}{\partial k^{\sigma}}A^{-1}]=
-\frac{i}{2(k^2)^2}\frac{2}{i}\int \frac{d^4p}{(2\pi)^4}\frac{1}{p^2-m^2}{\rm Tr}e^2[k^{\alpha}K_{\alpha}k^{\beta}K_{\beta}\gamma^{\rho}(\gamma^{\tau}p_{\tau}+m)\gamma^{\sigma}(\gamma^{\eta}p_{\eta}+m)]=
\nonumber\\
&&=\int d K_{\mu} -ig^{\rho\sigma}\frac{1}{k^2}\frac{2}{64\pi^2}[-\Lambda^2+4m^2\ln[\frac{\Lambda^2}{m^2}]]K_{\mu}^2e^2
\label{rez446}
\end{eqnarray}

We are forced to use a cut-off procedure along with an Euclidean space for both momenta and the field $K(k^2=0)$ that we denoted simply by $K_{\mu}$. Note that the actual result gets divided by the zeroth order partition function and all other factors dependent on the other $K_{\mu}(q)$ variables get canceled. Thus the result can be written as:
\begin{eqnarray}
 -e^2\frac{2}{64\pi^2}[\Lambda^2-4m^2\ln[\frac{\Lambda^2}{m^2}]]\frac{ \int d K_{\mu} K_{\mu}^2}{\int d K_{\mu}}=
 -e^2\frac{1}{16\pi^2}\frac{1}{3}[1-4\frac{m^2}{\Lambda^2}\ln[\frac{\Lambda^2}{m^2}]],
 \label{rez6789}
 \end{eqnarray}
where the integral over K is performed in spherical coordinates in the four dimensional Euclidean space with a cut-off $\Lambda$.  Here we need to note that the actual $K_{\mu}(p)$ coordinate has dimension of $m^{-3}$ but since it comes with an integral $d^4p$ one includes this factor in the variable to get a dimension of m. Then when one consider the traces  one needs to include a factor of $\frac{1}{V}$ to go from summation to integration. This  multiplies the result by an extra factor of V which is equivalent to dividing by $\frac{1}{\Lambda^4}$.

 The two loop term is calculated as easily:
 \begin{eqnarray}
&&\int d K_{\mu}i\xi\frac{1}{2}\det A{\rm Tr}[\frac{\partial A}{\partial k^{\rho}}A^{-1}]{\rm Tr}[\frac{\partial A}{\partial k^{\sigma}}A^{-1}]=
ig^{\rho\sigma}\frac{1}{k^2}\frac{1}{3}\frac{1}{64\pi^2}(-\Lambda^2+3m^2\ln[\frac{\Lambda^2}{m^2}])^2(K_{\mu}^2)^2
\label{rez65789}
\end{eqnarray}

 This again needs to be divided by the zeroth order partition function which leads to:
 \begin{eqnarray}
 -e^4\frac{1}{64\pi^2}(-1+3\frac{m^2}{\Lambda^2}\ln[\frac{\Lambda^2}{m^2}])^2\frac{\int d K_{\mu} (K_{\mu}^2)^2}{\int d K_{\mu}}=
 -e^4\frac{1}{64\pi^4}(-1+3\frac{m^2}{\Lambda^2}\ln[\frac{\Lambda^2}{m^2}])^2\frac{1}{2}.
 \label{rez67589}
 \end{eqnarray}

\section{Conclusions}

By adding up the results in Eqs. (\ref{rez65789}) and (\ref{rez67589}) we obtain for the full correction to the charge renormalization the quantity:
\begin{eqnarray}
\Pi(0)=e^2\frac{1}{16\pi^2}\frac{1}{3}[-1+\frac{4m^2}{\Lambda^2}\ln[\frac{\Lambda^2}{m^2}]]-
e^4\frac{1}{128\pi^4}(\frac{1}{3}-2\frac{m^2}{\Lambda^2}\ln[\frac{\Lambda^2}{m^2}]+3\frac{m^4}{\Lambda^4}(\ln[\frac{\Lambda^2}{m^2}])^2].
\label{finalrez567}
\end{eqnarray}

 The structure that we obtain is not unusual since one expects that a naive cut-off procedure violates the Ward identity and leads to an infinite photon mass. However when we integrated only over the modes $K_{\mu}$ with $k^2=0$ we made the underlying assumption that the lower modes are relevant such that the cut-off should not be too high.
Also one cannot take the cut-off so low as the order of m because then the computation of the beta function does not make sense. We thus will consider $\Lambda>m$ but not $\Lambda\gg m$.
In this context it make sense to use the expansion $\frac{m^2}{\Lambda^2}=\exp[-\ln[\frac{\Lambda^2}{m^2}]]=1-\ln[\frac{\Lambda^2}{m^2}]+..$.  We claim that since this procedure contains implicitly the higher order loops it is natural to have higher power of logarithms even at two loops.

Finally one can compute from Eq. (\ref{finalrez567}) and obtain for the full beta function:
\begin{eqnarray}
\beta(\alpha)=\frac{\partial (\frac{\alpha}{\pi})}{\partial \ln[M^2]}=\frac{1}{3}(\frac{\alpha}{\pi})^2+\frac{1}{4}(\frac{\alpha}{\pi})^4,
\label{rez9899}
\end{eqnarray}
which agrees with the standard result for the two coefficients of the beta function that are renormalization scheme independent. (Here we replaced the cut-off scale by a renormalization scale M).

There are no higher corrections in this procedure to the result in Eq. (\ref{rez9899}).

\section*{Acknowledgments} \vskip -.5cm
We are happy to thank A. Fariborz for valuable comments about the manuscript.
The work of R. J. was supported by a grant of the Ministry of National Education, CNCS-UEFISCDI, project number PN-II-ID-PCE-2012-4-0078.

\end{document}